%% file: paper_iot_Energy_DA_Survey.tex
\documentclass{elsarticle}
\usepackage[ruled]{algorithm2e}

\SetAlFnt{\small}
\SetAlCapFnt{\small}
\SetAlCapNameFnt{\small}
\SetAlCapHSkip{0pt}
\IncMargin{-\parindent}

\usepackage{flushend}
\usepackage{url}
\usepackage{pgfplotstable}
\usepackage{pdflscape}
\usepackage{siunitx}
\usepackage{longtable}
\usepackage{tabu}
\usepackage{cleveref}

\usepackage{floatrow}
\newfloatcommand{capbtabbox}{table}[][\FBwidth]
\usepackage{url}
\usepackage[utf8]{inputenc}
\usepackage{amssymb}
\usepackage{amsmath}
\usepackage{subfigure}
\usepackage{cmap}
\usepackage[T1]{fontenc}
\usepackage[switch]{lineno}

\begin{document}
\markboth{M. Schmidt, C. {\AA}hlund}{Smart Buildings as Cyber-Physical Systems: Data-Driven Predictive Control Strategies for Energy Efficiency}

\makeatletter
\g@addto@macro\endfrontmatter{\enlargethispage{-2\baselineskip}}
\makeatother

%

\title{Smart Buildings as Cyber-Physical Systems: \\Data-Driven Predictive Control Strategies for Energy Efficiency}
%
%
%

\author[NEC,LTU]{Mischa Schmidt\corref{cor1}}
\ead{mischa.schmidt@neclab.eu}
\author[LTU]{Christer {\AA}hlund} 
\ead{christer.ahlund@ltu.se}

\cortext[cor1]{corresponding author} 
\address[NEC]{NEC Laboratories Europe, Kurf\"ursten-Anlage 36, 69115 Heidelberg, Germany.}
\address[LTU]{Dept.~of Computer Science, Lule\aa~University of Technology, Forskargatan 1, 931 87 Skellefte\aa, Sweden.}

\begin{abstract}
Due to its significant contribution to global energy usage and the associated greenhouse gas emissions, existing building stock's energy efficiency must improve. 
Predictive building control promises to contribute to that by increasing the efficiency of building operations. Predictive control complements other means to increase performance such as refurbishments as well as modernizations of systems.
This survey reviews recent works and contextualizes these with the current state of the art of interrelated topics in data handling, building automation, distributed control, and semantics. The comprehensive overview leads to seven research questions guiding future research directions.
\end{abstract}




\maketitle


%

\input{sec_intro}

\input{sec_background}

\input{sec_BuildingsEE}
\input{sec_conclusion}

\section*{References}
\bibliographystyle{IEEEtran}
\bibliography{sigproc}

%







\end{document}

%% file: sec_intro.tex
\section{Introduction}
\subsection{Motivation}

This survey focuses on the topic of energy efficiency in buildings by improving operations using information and communications technologies (ICT). That approach is complementary to building stock refurbishment and modernization. In 2010, buildings accounted for 41\% of the primary energy use of the US with close to 75\% of this consumption being served by fossil fuels \cite{BuildingsEnergyBook2011}. In the EU-28, fossil fuels are also responsible for almost 75\%  of the total energy consumption \cite{EUROSTAT_Energy}. 
Buildings used more than two-thirds of their energy consumed for space heating (37\%), water heating (12\%), space cooling (10\%) and lighting (9\%) purposes \cite{BuildingsEnergyBook2011}. For the US, residential buildings used slightly more than half of the total building energy consumption \cite{Allouhi2015118}. \cite{NYCgreenhouse2014} shows that in 2013, 60\% of New York City's emissions stemmed from buildings in general - residential buildings as the largest source accounted for 37\%. In Europe, the ODYSEE and MURE databases indicate that buildings accounted for 40\% of the EU-28 final energy use in 2012, with residential buildings being responsible for two-thirds of the total building consumption \cite{ODYSEE_MURE_Energy}. 
Various building life-cycle analysis (LCA) case studies reveal that for typical buildings, irrespective of the type of construction, the building operational phase "dominates the life cycle energy use, life cycle CO2 emissions" \cite{Chau2015395}. For conventional buildings, the operational phase accounts for up to 90\% of the life cycle energy consumption, for low energy buildings up to 50\%. These figures confirm earlier findings in \cite{Ramesh20101592}.


\subsection{Political Context and Survey Focus}
Recognizing the contribution of human-made greenhouse gases (GHG) to climate change, the 2015 UN conference on climate change held in Paris, France, developed an unprecedented climate framework agreement that was signed by 180 countries. The UN agreement marks a significant step towards globally coordinated efforts to reduce humankind's contribution to climate change \cite{COP21_Agreement}. Even before that agreement, e.g.~the EU issued its energy efficiency guideline 2012/27/EU \cite{Directive2012/27/EU} which requires 20\% savings on primary energy usage by 2020 and 50\% savings by 2050, compared to 2008. These targets translate into annual savings of 1.5\% for all EU member states. In 2016 more than the required minimum of 55 individual nations that jointly account for at least 55\% of GHG emissions formally ratified the agreement, including China and the US. Now being in effect, each country ratifying this agreement will develop individual action plans that detail how it intends to reduce its GHG emissions. The agreement formulates the aim to keep global warming below $2K$ compared to pre-industrial temperature levels - ideally even keeping warming below $1.5K$. Regularly, each country will report its progress on these plans, and will also develop further plan amendments. 

As buildings account for a major fraction of the total energy consumption politics aim to improve buildings' energy efficiency levels by issuing appropriate regulations. Typically, these rules target newly constructed buildings or modernization measures. Building labels and certifications such as EPBD (EU) and LEED (US) do value the presence of building automation systems positively. However, macroscopic works targeting building stock energy efficiency such as \cite{Allouhi2015118, Berardi2016, Balaras20071298, Harvey2009, SwingGustafsson2016202, Tettey2016132} do not explicitly discuss energy efficiency potential in light of the possibilities offered by predictive techniques as surveyed in Section \ref{Sec:BuildingEnergyPerformance}. This survey specifically targets buildings equipped with some level of building instrumentation and with sensors installed at strategic points to improve the efficiency of building operation - the lion's share of lifetime building energy use. It reviews recent studies that apply computational methods to implement predictive control strategies integrated into the daily building operation.  
Efficiency gains by these predictive methods are complementary to possible modernization and refurbishment measures. The surveyed works lead to research questions to guide future advances in this field. Other approaches rooted in analyzing building data, e.g.~along the lines of \cite{Xiao2014109}, which analyzes data offline in regular intervals to infer operational inefficiencies and enable building staff to adapt operation schemes manually, are not covered by this survey. Similarly, studies that purely focus on improving modeling accuracies such as \cite{Chou2016751,Zhang201694,HU2017} are beyond its scope - while they may become relevant as tools in predictive control work, there is no energetic impact by these studies per se.

\subsection{Structure}
This work is structured as follows: Section \ref{sec:back} provides background information to contextualize Section \ref{Sec:BuildingEnergyPerformance}, which summarizes recent literature on data-driven predictive control applications for buildings. Section \ref{sec:open} formulates open questions guiding future research. Section \ref{sec:conclusion} summarizes and concludes the survey. 

%% file: sec_background.tex
\section{State of the Art in Buildings as Green Cyber-Physical Systems}
\label{sec:back}
\subsection{Building Energy Application Key Performance Indicators} 
For any building energy application to act sensibly, appropriate key performance indicator (KPI) definitions are required. In an attempt to allow benchmarking of Energy Service Company (ESCo) efficiency measures and service contracting, \cite{C21D22} defined several KPIs of relevance, among which:

\begin{itemize}
\item \emph{CO$_2$ emissions.} Reducing these emissions is an intuitive target for buildings, considering the discussions surrounding GHG emissions. However, in the building domain, CO$_2$ is only measured for monitoring Indoor Air Quality (IAQ, see below), not in the context of the energy supply. Therefore, this KPI is usually derived from the energy consumption by a conversion factor related to the energy source as e.g.~provided in \cite{energiedatengesamt}. 

\item \emph{Comfort}: This term expresses how well a control application can create conditions in which human occupants feel comfortable. As this concept is very generic, the literature covers several different aspects:

For \emph{thermal} comfort, the current practice typically treats maintaining indoor air temperature or operative temperature ranges \cite{6153586,Oldewurtel201215} as a proxy to meeting comfort targets. However, these parameters do not reflect the actual thermal sensation of an individual due to a set of other factors, such as solar radiation or humidity \cite{ISO7730_2005}. 
For example, the solar radiation effect on comfort has been studied in \cite{marino2015mapping} and the references within. Industry and research communities express the need for appropriate thermal comfort definitions for the purpose of building control \cite{mavrik2011advanced, cigler2012optimization,Klauco2014explicit}. To date, the most common thermal comfort index adopted by international standards is Fanger's \emph{Predictive Mean Vote} (PMV) model \cite{fanger1970thermal}: ISO 7730 \cite{ISO7730_2005}, and the adaptive standards EN 15251 \cite{CEN15251} and ASHRAE 55 \cite{ASHRAE-55} rely on it. Derived from PMV, the \emph{Predicted Percentage Dissatisfied} (PPD) expresses dissatisfaction of occupants due to poor thermal comfort. The suitability of these comfort indexes and standards is subject to debate: studying classroom thermal environments \cite{Zomorodian2016895} concludes that these indexes are "mainly found to be inappropriate for the assessment". 
To overcome questionnaire-based methods traditionally used to assess thermal discomfort, \cite{Gauthier201670} investigates an alternative form of data collection in addition to temperature sensors: by observing occupants’ activities (e.g.~activating heating, pouring a hot drink, changing clothing level). 

Despite thermal sensation also \emph{Indoor Air Quality} (IAQ) can be a source of (dis-)comfort. CO$_2$ and humidity levels, as well as the concentration of different pollutants, are the main parameters of concern. For example, \cite{Balvis201660} models the impact of air conditioning in an office room in Panama City from measured temperature, humidity, and CO$_2$ levels. \cite{24793} provides a more extensive discussion of air quality and thermal comfort. 

Discomfort has a substantial socio-economic impact: based on the data of 3766 pupils taught in more than 150 different classrooms of 27 schools, \cite{Barrett2015118} identified a significant impact of the environmental factors light, sound levels, IAQ, and temperature on the academic progress. \cite{Lamb2016104} used online surveys to analyze the self-reported work performance of 114 office workers over a period of 8 months about perceived thermal comfort, lighting comfort and noise of their offices. Discomfort in one or more of these factors acts as stress that reduces work performance by 2.4\%-14.8\%. For a more comprehensive overview, we refer to \cite{AlHorr2016369}, a recent survey on how building occupants' discomfort affects productivity.

\item \emph{Energy.} Measured during a period of concern, typically in kilowatt-hours [kWh], kilojoules [kJ] or tonnes of oil equivalent [Toe]. 
Depending on the context of comparison and benchmarking, often the energy consumption of a period (e.g.~one year) is normalized per visitor (e.g.~public buildings), employee (e.g.~office buildings), or floor area. When considering heating and cooling systems, weather normalization by Heating/Cooling Degree Days \cite{DegreeDays} is appropriate. That allows comparing consumption across climatic zones and years. As indicated in \cite{SwingGustafsson2016202}, there is a crucial difference of perspective between assessing energy efficiency from a \emph{primary} energy (i.e.~the total energy of the natural resource used) or a \emph{final} energy viewpoint (i.e.~the final use form, e.g.~used for electricity or space heating). Studies and surveys targeting political frameworks and policies usually reason about the primary energy effects whereas studies on building equipment or operation strategies typically take the final energy perspective. So-called primary energy factors (PEFs) establish a connection between both energy notions. However, there are variations in the definition and calculation of PEFs that can have significant consequences e.g.~when comparing different heating systems regarding primary energy use \cite{SwingGustafsson2016202}. 

\item \emph{Exergy} measures the maximum available energy for doing work. A thermodynamic system's exergy depends on the distance to the system's equilibrium. Unlike energy, exergy is not conserved. According to the Second Law of Thermodynamics, exergy is related to the quality and quantity of energy. Thus, a control scheme around exergy must address energy quality in addition to quantity \cite{Razmara2015555}. In theory, using boilers as the heat source in buildings creates a mismatch between exergy supply and demand, which should be avoided. For example, low-temperature floor heating outperforms other (high-temperature) space heating systems regarding exergy \cite{Kazanci2016119}.

\item \emph{Green factor.} The fraction of the energy used from renewable energy sources divided by the total energy consumption. 

\item \emph{Light} levels, measured in lux, are relevant for applications of smart blinds and lighting control, often as part of a comfort KPI assessment. 
 
\item \emph{Temperature}, measured in the controlled zone or system, can be used as an absolute reading or put in relation to an application specific target temperature. Often, temperature is part of comfort KPI assessment.

\item \emph{Underperformance Time (UPT).} Building systems have a defined range of indoor conditions related parameters, e.g.~temperature, CO$_2$ levels, relative humidity or lux levels. Often, this range may only be in effect during a distinct period, e.g.~office hours. UPT measures the time the system did not meet the target range when it should have.

\item \emph{Underperformance Ratio (UPR).} The UPT in relation to the amount of time the target range was in effect. 
\end{itemize}

\subsection{Buildings as Cyber-Physical Systems}
Following \cite{Lee:EECS-2008-8}, "Cyber-Physical Systems (CPS) are integrations of computation and physical processes. Embedded computers and networks monitor and control the physical processes, usually with feedback loops where physical processes affect computations and vice versa". Actions taken by CPS are not reversible \cite{DBLP:journals/itiis/GunesPGV14}. A delineation from the related field of \emph{Ambient Intelligence} \cite{Aarts2009} is that the physical processes controlled are not always subject to human interaction. The CPS concept as "co-engineered interacting networks of physical and computational components" contributes to advances in the field of Smart Buildings among others \cite{NIST-CPS}. Newly constructed as well as already pre-existing buildings are often already equipped to a certain degree with building automation infrastructure.  Typically, the current automation and control strategies are simplistic, e.g.~heating system supply temperatures being chosen based on current outside air temperature or system operation run based on fixed schedules. While the CPS definition of \cite{Lee:EECS-2008-8} allows for simple rule-based mechanisms, this work interprets CPS as using computational representations of the underlying physical processes to implement \emph{predictive} control strategies effectively. 

\begin{figure}[t]
\centering
\includegraphics[width=0.9\textwidth]{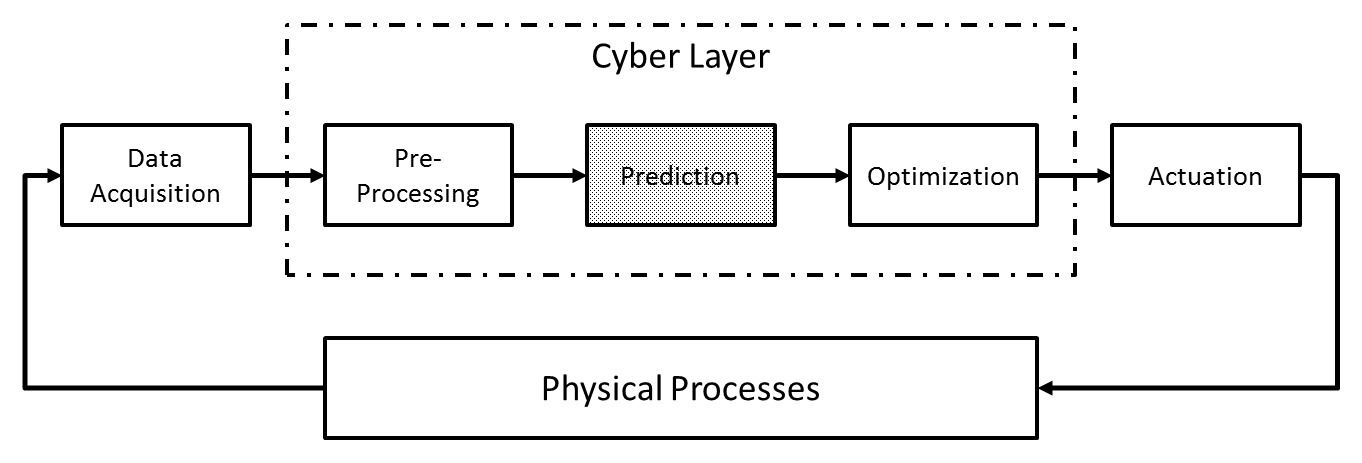}
\caption{Block diagram of the predictive optimization CPS control loop. Recent works use different types of approaches for the prediction step (shaded) as presented in Section \ref{Sec:BuildingEnergyPerformance}.}
\label{fig:cps-flow-loop}
\end{figure}

Figure \ref{fig:cps-flow-loop} illustrates the concept adopted in this survey:

\begin{enumerate}
\item Sensors and other information sources collect information on the building and its surroundings. This step is related to a number of fields: Wireless Sensor Networks (WSN) \cite{7145379}, the Internet of Things (IoT) \cite{7061425}, Machine-to-Machine communications (M2M) \cite{6674156, computers3040130, 6644332}, Sensor and Data Fusion for increased accuracy and temporal resolution \cite{7392809}, Pervasive Sensing as discussed in \cite{Kumar2016145}, and Building Automation \cite{Merz2009bas}. The scope of this survey is to use the information obtained for \emph{predictive} control. However, also \emph{reactive} strategies benefit from the information: e.g.~\cite{6740031} combines occupancy detection with schedule-based HVAC operation to increase energy savings.

\item To predict the evolution of the controlled building's physical processes within a defined time horizon computational representations of these processes are used. These predictions allow optimizing control decisions. Without loss of generality, this work builds on the usage of sensor data and other information sources in the optimization as well as in the continuous tuning of the representations. This step comprises multiple aspects:

\begin{enumerate}
\item \emph{Pre-processing}, converting, cleaning, selecting, and standardizing data. Many data-driven techniques are designed to operate on numeric features. This requires mapping categorical features to numeric values - a typical example is e.g.~the day of the week when integer values represent the weekdays. 
Cleaning numerical data from outliers with statistics-based data mining approaches as e.g.~documented in \cite{Xiao2014109, 6674155, tuprints3012} often helps to achieve satisfactory performance of the subsequent applications. The field of feature selection is a prominent research area surveyed in e.g.~\cite{Chandrashekar201416, 6918213, guyon2003introduction}. Regarding numerical stability, many techniques benefit from standardizing each input feature. There exist different procedures to standardize data, e.g.~shifting each feature datum by the feature's mean and dividing by its standard deviation. This approach centers each feature around the origin with a standard deviation of 1. Also, applying techniques such as Principal Component Analysis (PCA) \cite{jolliffe2002principal} and the field of representation learning \cite{6472238} can boost accuracy and prediction performance of the cyber-representations. On top of that, recent advances have shown the suitability of Deep Learning \cite{Schmidhuber201585} for "discovering intricate structures in high-dimensional data" \cite{LeCun2015} - which allows the unsupervised discovery of highly non-linear, abstract, and meaningful feature representations from training data. Building data usually accumulates as time series data, for which \cite{Langkvist201411, DEB2017902} provide overviews of different deep learning and more general machine learning approaches, respectively.  

\item Cyber-representation for \emph{prediction}. Section \ref{Sec:BuildingEnergyPerformance} outlines recent works categorized into two broad categories: \emph{theoretical approaches} and \emph{data-driven approaches}.  Over time, the accuracy of a cyber-representation may deteriorate as a building's environment is in constant flux. For example, occupants' preferences, as well as their behavioral and occupational patterns, change gradually or abruptly; building systems degrade, become repaired or replaced; refurbishments and modernizations take place; spaces are redecorated, and the weather changes with the seasons and among the years. Therefore, the representation requires regular updates to ensure satisfactory performance. For the theoretical approaches in Section \ref{Sec:BuildingEnergyPerformance}, the models need to be maintained by experts or, in case digital building models are used for building simulations, by computer tools. For the data-driven models, the field of \emph{concept drift}, surveyed in \cite{7296710, Gama:2014:SCD:2597757.2523813}, investigates the handling of changing environments in machine learning. 

\item \emph{Optimization}. A wide variety of well-known analytic optimization techniques can be applied, e.g.~linear programs, if a given building's problem formulation and cyber-representation are tractable. Considering that several stochastic events influence building operations in daily life, optimization is required to operate in an environment with aspects of uncertainty. For example, \cite{Sun2017} shows in the field of HVAC control several studies focused on improving control robustness by accounting for uncertainty. A more general review of research in the field of \emph{Robust Optimization} is provided by \cite{Gabrel2014471}. We refer to \cite{Gorissen2015124}, describing a practical guide on how to apply robust optimization to specific problems. 
When optimization problems become intractable, nature-inspired heuristics such as the following can be applied to find (nearly) optimal solutions. 

\begin{itemize}

\item \emph{Simulated Annealing} is a popular heuristic for optimization. Its primary operation consists of a local search to minimize a problem-specific cost function. As local search methods are prone to getting trapped in local optima, Simulated Annealing attempts to avoid entrapment in local optima by sometimes proposing a move to candidate solution that increases (worsens) the value of the cost function. A configurable acceptance probability determines the acceptance or rejection of this uphill move. Focused on single objective optimization \cite{Junghans2015651} uses a Genetic Algorithm's solution (see below) as the initial parameter configuration of a simulated annealing algorithm modified to avoid uphill exploration. Then it applies the proposed hybrid optimization scheme to a facade optimization planning problem in different climates validated by a building simulation. The case studies in \cite{Junghans2015651} show that the combination both methods achieves robust optimization results. Further, their combination reduces computational complexity compared to a repetitive use of the Genetic Algorithm to verify the optimization outcome.
 
\item The \emph{Particle Swarm Optimization} (PSO) is another popular heuristic. It relies on a population (denoted \emph{swarm}) of candidate solutions (\emph{particles}). The heuristic is based on a gravitational metaphor to iteratively update the particles according to simple rules of attraction and inertia. Various variants and applications exist as illustrated in \cite{zhang2015comprehensive}, e.g.~\cite{Wei2015294} extends it with the ability to address multiple objectives by calculating the Pareto front of HVAC operation. This ability allows specifying a trade-off between saving energy and addressing comfort aspects. 

\item While many more nature-inspired heuristics for optimization exist, \cite{Weyland201597} argues most of these only differ marginally from PSO. For example, the \emph{Firefly Algorithm} \cite{yang2010nature} is for specific parameterizations equivalent to PSO. This algorithm, inspired by the flashing behavior of fireflies aiming to attract other fireflies (configurations of decision variables) by means of brightness (the cost function to be optimized), is applied in \cite{Zeng2015393} to optimize multi-zone HVAC operation in a dedicated HVAC test facility outperforming standard PSO. 
To improve the balance between exploration and exploitation \cite{Coelho2013273} modifies the Firefly Algorithm's attraction equation with Gaussian distributions to avoid premature convergence to local minima. 

\emph{Cuckoo Search} is another heuristic similar to PSO that identifies problem solutions with bird nests and decision variable selections as eggs within the nests \cite{5393690, Weyland201597}. The algorithm draws on cuckoos placing eggs at random in the nests and an evolutionary aspect in that the best nests (containing high-quality eggs) will carry over to the next generation. However, for bad nests, the host bird owning the nest may discover the cuckoo egg and throw it away. With a configurable probability, it may even abandon the entire nest and build a new nest. This concept is extended by \cite{Coelho2014237} with a differential approach that lends a mutation operation inspired by the Genetic Algorithm to avoid that local search gets stuck in local minima. \cite{Coelho2014237} applies the concept to slightly different case studies than \cite{Coelho2013273} but achieves qualitatively similar results.  

\item According to the survey \cite{Shaikh2014409}, the single most widely used nature-inspired heuristic in the building optimization field is the \emph{Genetic Algorithm} \cite{holland1975adaptation}. This heuristic is a stochastic technique inspired from genetic recombination found in the process of natural selection.  
This iterative approach mimics biological evolution, searching a population of candidate solutions (represented by \emph{chromosomes} consisting of \emph{genes} - choices for optimization variables) for its fittest members. A problem-specific cost function expresses this fitness, and during the evolutionary process only the fittest members’ genes are mutated and exchanged stochastically to improve the solution. For example, \cite{Ascione2016131} demonstrates its use for scheduling HVAC operation decisions: validation by simulation indicates system operation cost savings of 56\% and improvements in thermal comfort.

\end{itemize}
\end{enumerate}

\item Control decisions are communicated to the building infrastructure to steer the physical processes as desired. In this step, again IoT and M2M aspects apply. 
Potentially, the decisions may be in the form of set-points that are communicated to lower layer control loops of the building automation infrastructure. This case resembles the approach of \emph{supervisory control}, typically executed by experts supervising plant operations. 
The decisions, however, may also be directly communicated to actuators, which effectively constitutes a control loop.

\item Actuation impacts the physical process, affecting the sensor information after a process dependent time delay.

\end{enumerate}

\subsection{Relation to Building Automation}
In building automation, hierarchical system structures are very common, typically designed in a three-layered architecture \cite{Merz2009bas}:
\begin{enumerate}
\item The lowest layer, the so-called \emph{Field Level}, consists of sensors and actuation devices.
\item The middle layer (\emph{Automation Level}) consists of controllers implementing control loops to meet configured set-points. 
\item The top layer, the \emph{Management Level}, usually consists of the computer hosting the Building Management System that offers a user interface and allows configuring static set-points as well as rules and schedules to change these set-points.
\end{enumerate}

\begin{figure}[t]
\centering
\includegraphics[width=0.8\textwidth]{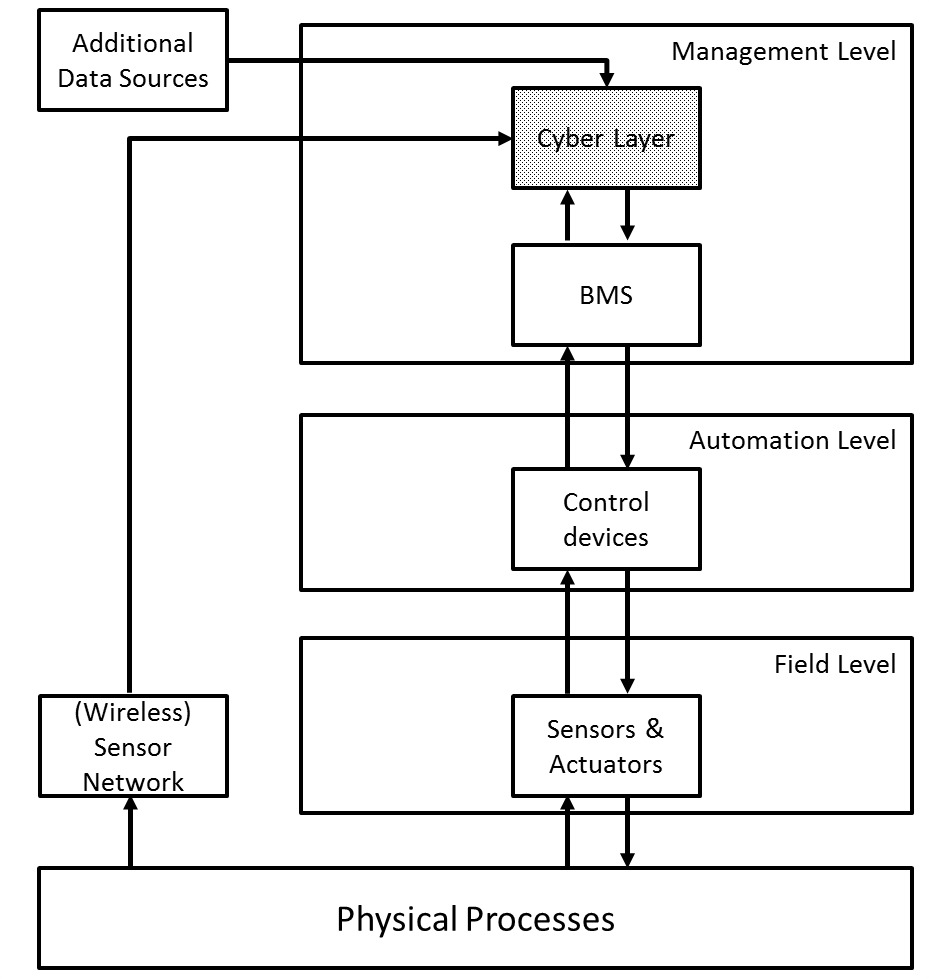}
\caption{Integration of the predictive CPS concepts of Figure \ref{fig:cps-flow-loop} (shaded) with the common three layer BMS structure described in \cite{Merz2009bas}.}
\label{fig:bms-cps}
\end{figure}

Communication protocols encountered at the different levels are M-Bus, Modbus, BACnet, EIB/KNX, LON and more recently also OPC.
Traditional building automation systems that follow this structure are reactive Cyber-Physical Systems. The delineation to this survey is the aim to improve building control by \emph{predictive actions}, e.g.~by appropriate set-point manipulation to address anticipated situations ahead of time. The advocated approach of operating buildings with predictive control strategies benefits from integrating pre-existing building automation infrastructure. Ideally, the BMS supports this by acting as a single gateway to the automation infrastructure enabled by the protocols mentioned above. Depending on the cyber-representation and the optimization goals, additional field level and automation level devices may be necessary. Also, the integration of other data sources e.g.~provided via the Internet, may improve the efficiency of predictive control strategies. For example, \cite{Schmidt:2015:EEG:2821650.2821661} deployed its cyber-representation for predictive control actions on a separate server by using the BACnet/IP protocol for BMS integration and accessed additional weather information provided by \cite{wunderground}. Figure \ref{fig:bms-cps} illustrates this concept. 

\subsection{Control Mechanisms}
Cyber-Physical Systems require real-time control, traditionally implemented in one of the three forms \cite{DBLP:journals/itiis/GunesPGV14} ordered by increasing sophistication:
\begin{itemize}
\item open loop control: based on only the desired value as an input signal, the CPS calculates control actions but lacks a feedback mechanism;
\item feed-forward control: the CPS takes into account additional environmental information collected from sensors and decides on actuation commands based on the anticipated relation between the physical system and its environment;
\item closed loop or feedback-based control: the CPS receives as feedback the difference between the input and the output signals, enabling it to adjust its control decisions - both the physical and the cyber parts of the system affect each other. 
\end{itemize}

Assuming a smart building equipped with sensors and automation infrastructure, predictive control of a building is a prime example of a closed loop CPS: actuation decisions will be reflected in future building data, which has the potential to affect future actuation decisions.

\subsection{Distributed Building Control and Conflicts}
The interplay of various systems with individual characteristics characterizes building operations. Some of these systems have diametric effects, e.g.~heating and cooling. In large facilities, typically the coordination of these systems is configured by human experts based on their experience and operational know-how. The literature explores methods to optimize building performance according to KPIs while freeing staff from the need to monitor the routine operation closely. In principle, when moving to automated building control, a centrally formulated constrained optimization problem could accommodate and balance the systems’ effects.  However, for large facilities, this centralized approach is computationally complex. When facing such complex problems, it is common to split them into multiple smaller, easier to solve problem formulations. In these settings, the \emph{Multi-Agent System} (MAS) design paradigm is popular as it is scalable, distributed, and manageable \cite{Labeodan20151405,Moradi2016814}. Several surveys in the domains of power networks \cite{Moradi2016814}, microgrids \cite{Kantamneni2015192}, and smart buildings \cite{Labeodan20151405} underline the concept's adoption across sectors. The popularity of MAS in the building control community is evident in \cite{Shaikh2014409}. 

Agents are entities that autonomously interact with their environment (and other agents) according to agent-internal rules. 
When implementing an entire building control as a multi-agent system, conflicts of the agents' decisions may arise in certain situations: distributed control "contributes to the rise of organizational conflicts due to goal or perception difference between agents" \cite{Lee2010574}. 
Based on work on inter-personal conflict resolution \cite{JOB:JOB4030130307}, \cite{Carreira201437} categorizes the different types of conflict handling based on the specific agents' levels of assertiveness and willingness to cooperate: 
\begin{itemize}
\item Avoidance: when agents have low levels of assertiveness and willingness to cooperate
\item Accommodation: when agents have a low level of assertiveness but a high willingness to cooperate
\item Compromise: when agents have medium levels of assertiveness and willingness to cooperate
\item Competition: when agents have a high level of assertiveness and but a low willingness to cooperate
\item Cooperation: when agents have high levels of assertiveness and willingness to cooperate
\end{itemize}

The capability to resolve possible conflicts in distributed building control is key to efficient and reliable operation in everyday life. As identified by \cite{Labeodan20151405}, a survey of multi-agent systems in commercial and residential buildings, agents that use so-called \emph{Utility} functions can weigh among different conflicting options, the likelihood of success, and goal importance. However, to accommodate conflicts with other agents, additional means, such as dedicated mediating agents or negotiation protocols, are required. In general, \cite{Kantamneni2015192} categorizes MAS architectures in \emph{centralized}, \emph{distributed} and \emph{hierarchical} describing the interactions and the distribution of responsibilities among agents. Focused on the particular setting of distributed model predictive control (DMPC) \cite{Christofides201321} defines settings where agents optimize their individual cost functions as \emph{non-cooperative} and settings in which each agent optimizes a shared  cost function as \emph{cooperative}. 

\subsection{Smart Grid Interactions, Renewables, and Storage Capacity}
Existing buildings' service systems may receive energy from a variety of grids: electricity, gas, district heating, or district cooling grids feed buildings worldwide. Traditionally, grid infrastructure goes hand in hand with centralized generation and distribution, which suffers from energy losses along the distribution network and potentially from supply-demand imbalances. In particular, the electricity grid is recently evolving towards a decentralized architecture with an increasing amount of distributed generation sources. The decentralized feed-in of electricity has considerably increased in recent years due to the unbundling of the energy supply chain, the technological advancements of renewable energy sources (RES), and political support promoting renewable installation. Therefore, electricity distribution system operators (DSO) have to cope with bidirectional energy flows in their networks. Furthermore, the variability of wind and photovoltaics is de-coupled from energy demand, which causes a supply-demand-imbalance in the electricity network \cite{TREBOLLE2010222}. Therefore, DSOs have to manage the fluctuating energy supply, for example by shedding renewables during times of high production, increasing the electricity network's transmission, installing additional capacities of energy storage, and by managing demand intelligently. For the latter, Demand Response (DR) is actively pursued by grid operators \cite{NationalGrid}. DR requires communication capabilities between the electricity grid and the consumers \cite{CUI2017720}. DR may result in an increased energy demand to take up surplus energy from the grid due to high RES generation. However, that higher demand is ``green'' as it is associated with low GHG emissions and can serve time-flexible loads that otherwise would be served by fossil fuels. 

Buildings with the capacity to store energy, in batteries (or connected electric vehicles), in hot or cold water tanks, or in their thermal mass are particularly well suited to leverage this energy buffering capacity to flexibly manage their demand and offer DR potential to the grid. Section \ref{Sec:BuildingEnergyPerformance} discusses several studies that demonstrate the feasibility of DR in data-driven building control, such as \cite{DBLP:journals/corr/RuelensCQSBB15, 7479093}. Embedded in a bigger smart city context international big scale research projects \cite{REPLICATE, mysmartlife} currently investigate energy demand-side management as one aspect contributing to higher energy efficiency in addition to other interventions, such as increasing citizens' awareness of energy efficient behavior. Apart from DR settings, predicting a building's renewable energy generation (e.g., rooftop solar panels' power output) to increase local RES consumption within the building, i.e., avoiding feed-in into the electricity grid, is another, though related, goal of efficient building operation. In the context of this survey, the CPS approach to predictive building control benefits from combining buffering capacity, demand prediction, and DR signals to manage demand flexibility and increase building operation energy efficiency by deciding on optimal control strategies. 

\subsection{Building Information and Semantics}
Ontologies conceptualize domain knowledge by establishing semantic relationships between classes and their properties. This form of knowledge representation enables applications to apply logical reasoning and inference. Several data models and ontologies suitable to inform predictive building control applications exist:
\begin{itemize}
\item The \emph{Industry Foundation Classes} (IFC) describe building and construction industry data in an object oriented model defined by buildingSMART \cite{buildingSMART} and adopted by ISO \cite{ISO-IFC}. The IFC exchange format definitions suit various disciplines, such as the facility management, during the different life cycle phases of buildings. An XML schema definition, as well as an OWL ontology of the IFC standard, exist. 

\item \emph{Green Building XML} (gbXML) \cite{gbXML} facilitates the transfer of building information stored in CAD building information models, enabling interoperability between building design models and various engineering analysis tools and models. It covers the planning phase as well as systems operational data. \cite{Reinisch:2011:TEE:1945420.1945421} transforms gbXML into an ontology for use with semantic web technologies. 

\item With \cite{TS103264}, ETSI SmartM2M endorses the \emph{SAREF} ontology describing Smart Appliances in residential environments. While SAREF focuses on appliances, it also models devices that control building spaces such as windows or doors. To cover building construction related aspects, ETSI SmartM2M relies on the FIEMSER data model and ontology \cite{FIEMSER-D5,FIEMSER}. 

\item Avoiding to model building physics or construction data, Project Haystack \cite{haystack} defines an ontology focused purely on building system operational data.

\item Furthermore, the literature provides also a multitude of information models targeted at enabling smart home automation services, e.g.~DomoML \cite{edsoai.698285358.2005}, DogOnt \cite{Bonino2008}, and BOnSAI \cite{Stavropoulos:2012:BSB:2254129.2254166}. Several works extended these to address energetic aspects in the home: e.g.~\cite{Bonino20111392} attaches an energy profile ontology for devices to \cite{Bonino2008}. Beyond that, \cite{Kofler2012169} defines energy semantics in a multi-agent smart home automation system modeling that not only cover device energy consumption but also the home's energy supply facilities and energy sources.  However, these domotics information models, even with energetic extensions, are typically not used for predictive control scenarios but rather for convenience services.
\end{itemize}

The survey \cite{Volk2014109} summarizes 180 research publications in the building information model (BIM) field dating back until 2002. For the existing building stock, \cite{Volk2014109} shows that BIM still is rarely used due to the challenges of high effort, BIM maintenance, and the question of handling of uncertain data, objects, and relations. It identifies as one major challenge for research the BIM creation for already 	existing buildings. Inspired by \cite{bazjanac2008ifc}, one possible application of BIM is described in \cite{jeong2014translating}: a translation between BIM and Building Energy Modeling (BEM) using Modelica \cite{Modelica}, an object-oriented, equation-based language to model complex physical systems. The work demonstrates the feasibility of reusing original BIM data in a BEM simulation. Another useful application is e.g.~WSN topology planning \cite{7145379}, i.e.~supporting one particular aspect of the approach presented in Figure \ref{fig:bms-cps} that enriches the information available to the predictive control strategies.

%% file: sec_BuildingsEE.tex
\section{Building Energy Performance by Predictive Control}
\label{Sec:BuildingEnergyPerformance}

\subsection{Building Energy Modeling Surveys}
Building energy efficiency is a multi-disciplinary field of ongoing research since decades. Surveys \cite{Li2014517, Harish20161272} review literature modeling building energy behavior. Their focus lies mainly on documenting the different fields and approaches to modeling buildings and their critical components, i.e.~energy systems, for building control and operation. From \cite{Li2014517} it becomes evident that despite considerable savings of energy or cost are envisaged, these have been validated in building simulation only, not in controlled experiments in routine operation. While \cite{Harish20161272} does not provide any energy savings figures, it reviews modeling approaches documented in literature since the 1980s, as well as the relevant parameters considered, the methods of validation, and available building simulation software packages. 
Survey \cite{Shaikh2014409} analyzes 121 works on the optimization parameters of concern in the field of building energy and comfort control. Its focus lies on assessing the publication trends and the techniques employed, rather than quantifying the effects achieved. The survey provides the insight that most published studies are US based and that Model Predictive Control (MPC), MAS approaches, fuzzy control, and the Genetic Algorithm are commonly applied in studies.

\subsection{Classification of Studies}
The remainder of this section surveys recent works on data-driven control of buildings: Table \ref{tab:energyStudiesLong} documents the method of validation (simulation or experiments) as well as the effect sizes achieved. Table \ref{tab:abbreviations} provides a list of abbreviations. This sub-section provides the taxonomy used to categorize the different approaches.

\begin{itemize}
\item \emph{Theoretical approaches}. These are based on either the physics of buildings or suitable approximations thereof. Works of this category often apply building (energy) simulation programs such as \emph{Modelica} \cite{Modelica} or \emph{EnergyPlus} \cite{Energy+}. In the literature, these kinds of approaches are commonly referred to as \emph{white-box} or \emph{model-based} approaches. As this survey puts emphasis on predictive control aspects rather than the modeling, both terms are used synonymously in this survey.
\item \emph{Data-driven approaches} train linear or non-linear regression models such as Neural-Networks (NN) based on observed (BMS) data.  In the literature, these approaches are commonly referred to as \emph{black-box} approaches. Sometimes they are also called \emph{model-free} to express the absence of a theoretical model. If information on the problem nature, e.g.~time-scales, effect sizes, linearity or non-linearity of the regression problem are taken into account to improve the training of these data-driven approaches, it is common to denote these approaches as \emph{gray-box} approaches. 
\item Combinations of data-driven methods with theoretical models are referred to as \emph{hybrid approaches}. 
\item Studies that experiment with approaches of different categories are classified as \emph{mixed}. In these cases, we focus on the best performing approach. 
\end{itemize}

\begin{table}[!t]%
\small
\centering
\begin{tabular}{ll}
\hline
Abbreviation & Explanation\\
\hline
C & Comfort\\
Cl & Clustering\\ 
DBN & Deep Belief Network \\
DD & Category of data-driven approaches\\
HW & Hot Water \\
E & Energy\\
EnvCo & Environmental context, e.g~ambient air  \\
			& temperature, solar irradiation, wind speed \\
FC & Forecast information about features, e.g.\\
		& weather forecast\\
GH & Grass Heating\\
H & Heating \\
HDD(E) & Heating Degree Day Normalized Energy\\
HuCo & Human Context, e.g.~occupation \\
HVAC & Heating, Ventilation \& Air-Conditioning\\
Hy & Category of hybrid approaches\\
L & Lighting\\
Lecture & University/teaching environment \\
Med & Medical building, Hospital, Care center\\
Mix & Mixed studies\\
MPC & Model Predictive Control\\
NN & Neural Network\\
Office & Office Building \\
OpCo & Operational Context, e.g.~other Systems'\\
 & Operational Data\\
P & Power \\
PV/T & Photovoltaic-Thermal System \\
Res & Residential Building \\
RL & Reinforcement Learning\\
RT & Regression Tree(s)\\
S & Shading \\
Stadium & Sports Stadium \\
SysOp & Controlled System's Operational Data \\
t & Time\\
Test & Test Facility \\
Th & Category of theoretical approaches\\
UPR & Under Performance Ratio \\
UPT & Under Performance Time\\
VE & Validation by Experiment \\
VS & Validation by Simulation\\
W & Window\\
X & Exergy \\
Z & Target Zone parameters, e.g.~IAQ \\	
\hline\\
\end{tabular}
\caption{Abbreviations for Table \ref{tab:energyStudiesLong}.}
\label{tab:abbreviations}
\end{table}

\subsection{Theoretical}
\label{Sec:WB}
Focused on the day-ahead planning of hourly HVAC set-points, \cite{Ascione2016131} combines thermal building simulations in EnergyPlus \cite{Energy+} with a Genetic Algorithm implemented in Matlab to calculate the Pareto front of operation concerning energy and comfort (PPD). Considering a typical day of the heating season in Naples, Italy, the simulations demonstrate savings of heating cost of up to 56\% compared to a standard control strategy when target user comfort allows a maximum of 25\% PPD. This result still improves the maximum observed PPD by 8\%. When targeting higher PPD (20\%), energy savings amount to 42\%.

Optimizing HVAC scheduling with model predictive control for buildings with several hundreds of zones becomes computationally intractable. For this reason, \cite{Radhakrishnan201667} divides the problem into three smaller, but logically linked sub-problems: knowing their characteristics and target indoor temperatures, controllers for individual zones compute the minimum required energy demand and request this amount from a central scheduling instance. That takes the building's chillers' characteristics as well as information of the building's chiller sequencing to decide on providing cooling energy to the zones. For each zone, this allocated cooling energy is required to meet the demand requested by the zonal controllers, but may also exceed the demand to exploit beneficial chiller characteristics. This way the central scheduler decides the appropriate ventilation fan stages and damper positions (expressed by desired duct pressures) to distribute the cooling energy through the ventilation ducts. A building simulation demonstrates that the distributed approach is close to optimality for a small number of zones and is also able to scale to hundreds of zones. Compared to an often used standard pre-cooling strategy, the awareness of each zone's thermal demand saves 17\% energy. 

\cite{Schirrer201686} proposes a novel modular MPC approach that explicitly models non-linear building dynamics. The concept allows approximation of the globally nonlinear optimization problem to solve the decoupled sub-problems efficiently: first, based on measured room temperatures, TABS temperatures, and weather prediction a trajectory of temperature evolution of the building is estimated. Second, required thermal flows of the building are calculated to adjust temperatures to desired ranges. Third, corresponding set-points are derived. The approach's effectiveness is demonstrated in a co-simulation controlling an Austrian low energy office building's thermal flow. However, no energy savings are quantified.

Given a cost budget for energy and using a linear indoor temperature evolution model, \cite{7288444} proposes a Comfort Prioritizing Greedy (CPG) algorithm to schedule appliances under consideration of their nature ((non-) deferrable and (non-) interruptible)). Validation is simulation-based and focuses on a single day. The CPG algorithm performs "slightly better" than both the standard bin-packing algorithm as well as a linear program to optimize the scheduling.

Focused on minimizing exergy destruction by HVAC use, \cite{Razmara2015555} shows in simulations that exergy destruction and energy consumption are reduced by up to 22\% and 36\% compared to traditional control. Compared to an energy-focused MPC, the simulations show 4\% exergy savings and 12\% energy savings.

In a Demand-Response (DR) setting \cite{doi:10.1080/15325008.2013.862322} combines a building's maximum demand with DR incentives in a single equation, constrained by appliances' characteristics. The approach combines rolling optimization of a multi-hour prediction horizon with minute-based real-time control strategies. A fuzzy logic controller controls appliances focusing on cost reduction: simulations demonstrate cost reductions of 16\%-19\%. A lab experiment using Zigbee illustrates the operational potential. The same authors focus in \cite{6787100} on the real-time scheduling of appliances using the conditional value at risk metric from economics to express uncertainty. Fuzzy logic decides on participating in a given DR instance, computation time amounts to 30-40 seconds per 30 minute planning interval. The forecasting of conditions, e.g.~prices or PV generation, is considered given. Validation is based on simulating a single day, which indicates an electricity bill reduction by 18\%. 

Studying six small and medium-sized commercial buildings in the cities of Boston, Chicago, and Miami \cite{Li20161194} demonstrates energy cost savings of 20\%-60\% for HVAC operation on five days in August when compared to a pre-cooling night-setback strategy using three-year average weather data. The savings depend on the individual building and its location as well as the relative weight of comfort compared to energy cost. The study applies PSO to identify the Pareto front of PMV versus energy costs in EnergyPlus \cite{Energy+}.

\subsection{Data-driven}
\subsubsection{Neural Networks}
\cite{Zeng2015393} optimizes multi-zone HVAC operation using NN to predict room temperatures and energy consumption. The study uses relative room humidity and room temperature as input to minimize energy subject to comfort conditions by controlling the supply air's temperature static pressure set points. The work proposes to perform the optimization by a Firefly algorithm. Validation is purely computational, based on data of a single day for which the Firefly algorithm outperforms PSO. Energy savings range between 2\% (most strict comfort constraints) and 17\% (most relaxed). \cite{Wei2015294} extends the energy optimization to consider also indoor CO$_{2}$ levels.  Compared to seven other regression models, a NN Ensemble performed best. A modified PSO solves for Pareto-optimal solutions of IAQ, comfort, and energy consumption. Different weightings of these objectives create different Pareto-optimal trade-offs. Computational results on the recorded two week period indicate average estimated electricity savings of 12\%-17\%.
Both works use expert input and feature selection algorithms to reduce the several hundred parameters sampled in 1-minute intervals (and averaged per 30 minutes and 60 minutes respectively) but do not elaborate on this aspect.

As an evolution of \cite{Ferreira2012238} that trades off HVAC energy consumption and user comfort using NN and multi-objective optimization, \cite{Ruano2016145} builds a data-driven predictive model for HVAC operation to minimize economic cost while ensuring comfort. The approach takes into account indoor temperatures, schedule information, cost, and weather variables. Exploration of lag times, i.e.~the length of history to consider in the models, is explored by a search heuristic. The work is validated in three lecture rooms of the University of Algarve, Portugal, in several experiments spanning a period of two weeks in June 2015. The results show financial savings while spending more energy to ensure minimized comfort violation for the HVAC unit under control: "savings in the order of 50\% are to be expected". Unfortunately, the work does neither normalize for room characteristics nor weather. 

Studying a more exotic heating system, \cite{Schmidt:2015:EEG:2821650.2821661} demonstrated weather-normalized thermal energy savings of 56\% over a winter season operating a soccer stadium's grass heating system - a major heat sink. This study experiments with a variety of control heuristics, e.g.~simple statistics-based methods as well as more advanced NN to predict the soil temperature evolution - the latter achieving better results. Best results are obtained by extending the control concept with awareness of operational context, i.e.~the status of other heat consuming systems to avoid bottleneck situations. Notably this work is the only work applying methods of statistical inference, enabling it to quantify confidence intervals for the different strategies' savings.

Starting from a thermal building simulation, \cite{7317804} proposes - after a pre-processing stage of sensitivity analysis and PCA - to use NN to learn building behavior regarding energy and comfort subject to control actions. Genetic Algorithm-based optimization is then applied to derive building control rules to be stored in a knowledge base that a facility manager can choose from, e.g.~to strive for energy savings targets. The approach is validated using three months of simulation and two months of experiments for a care home in the Netherlands where heating supply, window opening, the degree of shading, and light levels are controllable. Weather-normalized energy savings amount to approximately 25\%.

\subsubsection{Reinforcement Learning}
Using Reinforcement Learning (RL) to optimize the economic aspects of operating electric water heaters, \cite{DBLP:journals/corr/RuelensCQSBB15} demonstrates in simulations 24\% savings for using day-head prices and 34\% savings for the imbalance prices (stemming from forecasting errors) compared to the default strategy. A 40-day lab experiment achieves 15\% cost savings. When excluding the algorithm's exploration phase, savings reach up to 28\% confirming the simulation results. Features reflected are the day of the week, the index of the quarter of the day, as well as 50 temperature sensors. An auto-encoder compresses the features to five dimensions. 

Using an auto-encoder to reduce state vector dimensionality, \cite{ruelens2015learning} demonstrates by simulation of winter and summer seasons to be able to save 4\%-11\% energy of a heat pump with a set-back strategy in two different buildings. As no thermal or physical modeling is involved, the approach is transferable to other buildings and other climatic zones without requiring extensive effort. 

\cite{Costanzo201681} uses an ensemble of 40 NN to assist batch RL in creating an efficient HVAC DR controller able to control on-off decisions. A simulation of 40 days with different temperature regimes validates the approach. After collecting 16 days' data, the inferred control policies are stable within 90\% of the mathematical optimum. A shorter experiment in a living lab validates the findings qualitatively.

For a Swiss low exergy residential building \cite{Yang2015577} controls the mass flow parameter through a photovoltaic-thermal array to improve power output. Over the course of three simulated years, 5-11\% power improvement is achieved when compared to a rule-based controller configured by domain experts.

Focused on lighting and blinds operation, \cite{Cheng201643} presents an intuitive human machine interface via which users can provide feedback on their individually perceived light level comfort to a q-learning based controller. That aims to minimize lighting and HVAC energy consumption while avoiding violations of users’ minimum/maximum light level constraints. The usability is validated by a trial with ten students and two office workers lasting five months. Energy savings of up to 10\% are derived by simulating cloudy and sunny conditions with different user preferences when comparing the reinforcement learning based controller with conventional automated lighting control.


\subsubsection{Regression Trees}
Motivated to maximize participation in DR programs, \cite{7479093} relies on regression tree ensembles to predict the building electricity baseline from environmental parameters and system state variables. By rearranging the regression trees into two stages where non-control parameters are represented at the top of the tree and control variables towards its bottom, a region of desired control variables can be inferred to derive DR strategy actions dynamically. The inference is executed every five minutes by a linear program. For a large reference building, assuming 20 DR events over summer, \$45,600 in revenue for participating in DR (37.9\% of the campus' energy bill) could be expected. The authors report a 17\% higher curtailment than a rule-based DR strategy while maintaining thermal comfort. 


\subsection{Hybrid}
\cite{7170745} combines a linear resistance-capacitance model with a NN. The work shows by simulation and experiment at the check-in hall of Adelaide airport's terminal T-1 that this hybrid approach successfully combines the non-linear approximation ability of NN with the capability to extrapolate to new situations. The approach learns optimal start-stop pre-cooling strategies for HVAC operation. Pre-cooling saved 13\% electricity cost at the expense of 5.6\% increased energy consumption, while the start-stop strategy during hours of occupation achieves up to 41\% energy savings.

\cite{Gruber2015934} uses a simplified model-based controller with non-linear filters for indoor climate control. The study experiments in a mock-up office and meeting room environment in winter and summer seasons in Sweden. By closely taking into account IAQ, the controller reduces room ventilation rates by 12\% - 19\%. 

\subsection{Mixed}
\cite{Baldi2015829} applies Parametrized Cognitive Adaptive Optimization (PCAO) to optimize a ten-office building in Greece. The work studies two variants: (simulation-)\emph{model-based} and \emph{model-free}. Before deploying a PCAO-derived controller to the real building, its performance is assessed by a thermal building simulation. The model-free approach directly applies a control mechanism to the building after estimating a so-called \emph{performance index}. In simulations, compared to two baseline rule-based controllers, model-based PCAO reaches energy savings of 45\% and 25.6\% respectively. At the same time, comfort is enhanced by 7.0\%-8.7\% and 30.7\%-33.5\%, respectively. In these simulations, model-free PCAO achieves energy savings 41.2\%-44.6\% and 20.5\%-25.3\%, respectively. Simultaneously, comfort improves by 3.5\%-7.8\% and 28.1\%-32.9\%. However, in real-life experiments on different days in 2012 and 2013, model-free PCAO achieves energy savings of 19\%, outperforming the model-based approach due to modeling inaccuracies.

\begin{landscape}

\begin{table}[!t]%
\footnotesize
\centering
\begin{tabular}{lcccccccccl}
\hline
Ref. 	&  Building  	&  Systems   &  BMS & Validation & Features  & Time res. &  Category & Conflict 	&  KPI & Effect size \\

			&  Type				&  optimized & integr.	 & Method \& Period		&	& Interval & 		&  handling 				& 	& \\

\hline

\cite{Razmara2015555} &  Lecture 	& HVAC  &  & VS:1D  	& EnvCo, HuCo, Z & 1h & Th & 	&  X	& -4\% X, -12\% E\\

\cite{Ascione2016131} & Residential & HVAC & & VS:1D & Cost, HuCo, Z & 1h & Th & & Cost, C & [-56,-42]\% Cost \\ 

\cite{Radhakrishnan201667} & & HVAC & & VS:1D & EnvCo, HuCo, SysOp, & 30m & Th & & E, C & within 2\% optimality, \\ 	
 &   &		&		&		&  Z &&&&&-17\% pre-cooling E\\

\cite{Schirrer201686} 	& Office 	& HVAC 		& & VS:1W & EnvCo,  FC, HuCo,   & 15-30m & Th & & E,C & \\ 
 & 	 &  	&  & 		 & SysOp, Z \\

\cite{7288444} & Res  & HVAC & 	 			& VS:1D 			& EnvCo, SysOp, Z   & 1h  & Th &  &  C & Max. Comfort\\
 &   &   &   & 		&  Price, FC \\

\cite{doi:10.1080/15325008.2013.862322} & Res & HEMS &   & VS:1D & EnvCo, HuCo, SysOp,  & 1m & Th  & &   Cost & [-19,-16]\% Cost \\
 &   &   &   &    & Price, FC &   &  &   &  & \\

\cite{6787100}  & Res & HEMS & & VS:1D & EnvCo, HuCo, SysOp,  & 30m & Th  & &   Cost & -18\% Cost \\
 &   &   &   &    & Price, FC &   &  &   &  & \\

\cite{Li20161194} & Office &  HVAC & & VS:5D & C, Cost, EnvCo, & 30m & Th & & Cost, C & [-60,-20]\% Cost \\ 
 & 	 &   &   &   &  HuCo \\

\hline

\cite{Wei2015294} & Test & HVAC &   & VS:2W	& EnvCo, SysOp, Z	& 60m &  DD:NN &  &    E, C & [-17,-12]\% E \\

\cite{Zeng2015393} 		  & Test & HVAC &   & VS:1D 	& EnvCo, SysOp, Z  & 30m &  DD:NN &  &   E, C & [-17,-2]\% E \\

\cite{Schmidt:2015:EEG:2821650.2821661} &  Stadium & GH &  $\checkmark$ & VE:3M &  EnvCo, FC, OpCo,   & 10m &  DD:NN & $\checkmark$ & E,C & -56\% norm. E \\
 &  &  &    &  &   SysCo, Z &  \\

\cite{Ruano2016145} &  Lecture 		& HVAC 	& $\checkmark$ 	& VE:2W 				& EnvCo, FC				& 5m				& DD:NN 	&  &  E, C & -50\% Cost expected\\

\cite{7317804}  & Med& H,L,S,W & $\checkmark$ & VS:3M,VE:2M & EnvCo, HuCo, SysOp, & 15m & DD:NN & & E, C & -25\% norm.~E\\ 
 & & & & &  Z \\

\cite{DBLP:journals/corr/RuelensCQSBB15} &  Test & HW &  & VE: 40D & FC, Price, SysOp,   & ? & DD:RL &  &  Cost & [-34,-15]\% Cost \\		
 & & & & & t \\

\cite{ruelens2015learning} & Res & HP &   & VS:1 su.,1 wi. & EnvCo, FC, SysOp,  			& 15m & DD:RL &  &  E & [-11,-4]\% E \\		
 & & & & & t \\

\cite{Costanzo201681} & Test & HVAC & & VS:40D,VE:40D  & EnvCo, FC,	Price,  & 5m   & DD:RL &  &  Cost, C & within 90\% of optimum \\		
 &  &  &  &   &  SysOp, t, Z \\

\cite{Yang2015577} &Res  & PV/T 	&   & VS:3Y 					& EnvCo, SysOp   & 30m   & DD:RL 	& 	&  E 	& [6,11]\% P output\\

\cite{Cheng201643}  & Office & L, HVAC & & VS:2D & E, EnvCo, Lux  & ? & DD:RL & & E, C & -10\% E\\

\cite{7479093} & various & various &  & VS:4M & EnvCo, FC, HuCo,   & 5m  & DD:RT &  &  E & +17\% curtailment \\
 &   &   &  &   & SysOp, t, Z \\

\hline 

\cite{7170745}  & Airport & HVAC & $\checkmark$  & VS:1D,VE:4D & EnvCo, FC, HuCo,  & 10m & Hy &  & E, C, Cost & Start-Stop -41\% E; \\
	&	 &	 &	 & 	&	Price, Z &&&&&Pre-cooling -13\% Cost\\

\cite{Gruber2015934} & Test & HVAC &   & VE:2D &  EnvCo, HuCo, SysOp,  & ? &   Hy & &  Flow & [-19,-12]\% Flow \\
	&  &   &   &  &  Z &   &   & &   &  \\

\hline

\cite{Baldi2015829} & Office & HVAC & $\checkmark$ & VS:2W,VE:2W &  EnvCo, HuCo, SysOp, & 10m & Mix & &  E, C & VE: -19\% E \\
	& & & & & Z   &&&&& VS: [-25.3,-20.5]\% E, \\
	& & & & &    &&&&& [3.5,7.8]\% C\\

\end{tabular}
\caption{Categorization of studies using data-driven predictive control for improving building energy efficiency. Table \ref{tab:abbreviations} explains the abbreviations.}
\label{tab:energyStudiesLong}
\end{table}

\end{landscape}

%% file: sec_conclusion.tex
\section{Open Research Questions}
\label{sec:open}

In light of the reviewed research studies documented in Section \ref{Sec:BuildingEnergyPerformance}, this section identifies seven open research questions.

\subsection{Buildings as CPS - Methodology}
The simulations and experiments described in Section \ref{Sec:BuildingEnergyPerformance} demonstrate the positive effects of applying predictive control strategies to buildings. That promotes the widespread adoption of predictive control in buildings, to meet global greenhouse gas emission targets. However, the field lacks a concise methodology to develop and deploy highly effective strategies to specific existing buildings. None of the works in Table \ref{tab:energyStudiesLong} describes a general methodology for turning existing buildings into a closed-loop CPS supporting the deployment of predictive control strategies. 

The methodology to be developed could e.g.~be based on the established general Model-Based Design Methodology for CPS \cite{5982785} but adapted to the specifics of the building community to increase usability. For example, the methodology needs to account for the specifics of existing buildings, e.g.~to integrate already pre-existing automation infrastructure and building instrumentation. Further, especially in bigger commercial buildings, the methodology also needs to address the possibly complex stakeholder landscape. The potential is enormous: of all commercial buildings in the U.S., 42\% are equipped with automation systems \cite{BuildingsEnergyBook2011}. 

\emph{RQ 1: What is a suitable methodology to evolve existing buildings into a CPS for higher levels of operational efficiency?}

\subsection{Feature Selection}
Selecting meaningful features is a cornerstone to creating well-performing predictive models. In the SCADA of a common medium or large-scale building, a considerable number of variables is available for study. For example, in \cite{Schmidt:2015:EEG:2821650.2821661} the soccer stadium's BMS provides access to approximately 13,000 variables. The binary question whether or not to include a feature in combination with other variables yields $2^N$ possible combinations. The number of possible combinations increases even more, when 
\begin{itemize}
\item studying the effects of varying the amount of history information for each of the variables considered;
\item applying different linear and non-linear techniques to transform or normalize the variables; 
\item applying different linear and non-linear modeling techniques; 
\item tuning the techniques’ hyper-parameters (e.g. number of hidden layers and neurons in a NN).
\end{itemize}
Discretizing the space of possible combinations to explore (often referred to as \emph{grid search}) reduces that number, but this heuristic has large effects on the optimality of the final result and the efficiency of building the predictive model. 

Unfortunately, the works in Table \ref{tab:energyStudiesLong} do not describe the rationale how relevant features (variables as well as their history lengths) are selected. While the field of feature selection is a prominent research area, additional expert knowledge can guide the feature selection process to reduce the overwhelming complexity of an exhaustive feature space search. As BIM is becoming prevalent in the building domain, its use to effectively select features in an algorithmic way should be investigated. While the notion of semantic feature selection itself is not novel, it has never been used for predictive building control applications. \cite{7317804} identifies relevant environment parameters and set-points from an ontological building description to derive control rules; however, the feature definition, as well as the relevant system or building operational context, are not studied. More generally, semantic feature extraction has been applied successfully in the areas of mining text \cite{989501} and image data \cite{6166371, Guo2007}.  Also, graph-based feature selection has been used  successfully in DNA sequencing \cite{yang2012feature} (available e.g.~in R library FGSG \cite{FGSG}). Further, a recent publication in the industry automation domain describes how a manufacturing process ontology can be processed to select meaningful features from a SCADA system on which then machine learning methods can be applied \cite{Ringsquandl2015}, but does not describe how appropriate history lengths for each feature can be deduced. 

To conclude, the concept of semantic feature selection should be extended to also take BIM information into account when considering the relevant history lengths for the different variables considered. For example, the thermal interdependencies between building zones could be derived from a wall's thermal admittance \cite{ISO-Admittance} and its materials. These parameters determine a wall's decrement delay (time lag between the timing of the temperature peaks at either of the wall's sides) and its decrement factor (dampening the temperature peak of one side of the wall to the other) \cite{Shaik2016}. That information should lead to the appropriate history lengths and inter-feature lags. 

Beyond studying smart feature selection based on BIM, it is evident that additional sources of information that describe the human context (e.g.~occupation schedules) and the environmental context (e.g.~weather forecast) need to be taken into account. In this line of thought, the BIM or another linked ontological information source could evolve to indicate connectivity to these information sources such as calendars or scheduling information.

\emph{RQ 2: How to perform semantic feature selection for predictive building control?}

\subsection{Inter-building Transfer}
The large number of buildings built anywhere on the planet renders savings by predictive control strategies impossible if they cannot be rolled out and adapted to new buildings efficiently. Commercial M2M platforms facilitate deployment of predictive control applications to new buildings by abstracting from lower layer communication details \cite{6644332}. Still, the transfer of a predictive control application to another building remains time-consuming, error-prone, and costly, because
\begin{itemize}
\item an adaptation of the variable names is typically necessary, i.e.~input and actuator data point names need to be mapped, and
\item the new building has different characteristics, i.e.~the predictive models may have poor performance and may require re-training with different features.
\end{itemize}

It is necessary to study effective approaches to address this. Conceptually we see the following fields as promising:
\begin{enumerate}
\item The transfer of a predictive control application to a new building can be seen as an abrupt and drastic form of \emph{concept drift}.
\item \emph{Transfer learning} focuses on solving new but similar problems by utilizing previously acquired knowledge. Typically, the feature spaces of source and target domain are assumed to be equal \cite{5288526,Lu201514}. 
\end{enumerate}

These fields may be complemented by approaches such \cite{Bhattacharya:2015:AMC:2821650.2821667, Balaji:2015:ZOL:2821650.2821674} that analyze the meta-information in a BMS, e.g.~the data point names as well as the data, by data mining techniques to identify and map variables of interest correctly with minimal human intervention. Further, e.g.~\cite{Cui2016251} could be used in combination with e.g.~transfer learning as it recommends the most appropriate Building energy model to use for a particular building - to which then transfer learning principles could be applied to map to the new building. 

 
\emph{RQ 3: How to transfer a particular building’s predictive control strategies to another building?}

\subsection{Control Conflicts}

Several human-centric examples of control conflicts stemming from users’ preferences and activities are documented in the literature on domotics and smart homes.  For example, \cite{Camacho20146161,RuiJoseLobatoCamacho__} introduce an ontology-based reasoning to automatically detect and resolve conflicts stemming from different users' requirements. Upon identifying conflicts on environment variables and activities, the works propose to use constraint programming to settle the conflict situation. The works use ontological reasoning to maximize user comfort and minimize energy consumption, but only provide qualitative, not quantitative statements on energy savings. \cite{doi:10.1080/10789669.2014.980683} uses agents to communicate in a smart building to reflect a user's personal preferences regarding Fanger's PMV to reflect this user's personal comfort appropriately. If multiple users are present, compromise preferences are looked for (and users might be suggested to change their clothes). \cite{Carreira201437} detects conflicts as changes of the environment state resulting in an undesired context of the application or user expressed by a constraint satisfaction problem. Conflicts are reasoned about in an attempt to find a compromise solution. \cite{6560391} introduces an ontology based multi-agent home automation system and resolves conflicts identified during the concept covering phase by a utility based negotiation scheme. The agents negotiate based their individually configured utility definitions that they keep secret. A central home automation mediator strives to optimize global utility. 

While these solutions address conflicting control decisions originated from human preferences and activities, other sources for conflicting control decisions are e.g. 
\begin{itemize}
\item resource scarcity, e.g.~due to coincidence factor based boiler capacity planning: the coincidence factor captures the amount of peak demand overlap; if all thermal building systems have their peak at the same time (i.e.~coincidence factor was chosen wrong or control strategies have changed), the boiler capacity will be insufficient. However, current works in buildings typically do not resolve shares/demands/capacity constraints in case of resource conflicts \cite{Carreira201437}: for example, \cite{4594887} uses priority-based queuing of user requests for resources and exclusively grants access to these. 
\item inter-dependent zones are heated and cooled at the same time, causing a potential waste of energy may result in a negative feedback loop of successively stronger control actions of the involved systems' control agents.
\end{itemize}


According to Table \ref{tab:energyStudiesLong}, the literature focuses on controlling a single building system serving one or more building zones. It seems feasible to encapsulate the individual system control approaches in a multi-agent system. However, only a single study reflects the impact of its control scheme onto other building systems and possible conflicts. It does so in a reactive manner by monitoring the controlled system's operational context. Predicting situations of conflict and scarcity is for further study. 

For situations of conflict in multi-agent systems, literature often assumes cooperation among agents towards a common goal. For settings of \emph{cooperative} agents where task planning is admissible, \cite{pecora2007multi} describes in the context of an Ambient Assisted Living application a coordinated multi-agent action scheduling subject to resource constraints that supports conflict resolution.

However, the data-driven predictive building control domain may also face scenarios of competitive agents. For example, complex multi-tenant structures within a multi-story office building may lead to dedicated agents per tenant. Capacity constrained resources, e.g. a limited boiler capacity, a shared PV installation providing cheap electricity, or participation in Demand Response (DR) events may plausibly be argued to cause competition among different control agents. When extending the control scenario to multiple buildings, the problem space becomes even more complicated: will multi-tenant houses cooperate to compete against other houses for larger shares of a neighborhood-level electrical storage, but compete for PV usage? The insight that tenants inside a building may rent multiple zones adds further complexities: assuming each zone is controlled by an agent, the tenant’s agents should intuitively be willing to cooperate, while they might be in competition with other tenants’ agents. As tenant structures may change over time, as well as tenants' attitudes and policies towards each other, coalitions of cooperative agents of different tenants may emerge and disappear over time. The insights of \cite{bikakis2014computing, scully2014forming} may help address this setting. However, agents need to be aware of their own and the other agents' association. A possible way to realize this is enhancing agents with self-descriptiveness as introduced in \cite{fahndrich2013self, Fahndrich2013}.   

In \emph{competitive} settings, possibly issues of trust and data access need to be resolved or negotiated. Either, agents directly get access to the variables relevant for their operational decisions, or, if e.g.~access on another agents' variables is an issue, appropriate communication schemes and the identification of the correct responsible agent are also issues. However, some variables of the operational context may be less controversial to share among agents of different tenants: for example, the room temperatures of workers' offices may be more contentious to share with another tenant's agent than the supply temperatures and the working status of the ventilation. In certain competitive situations, a dedicated agent representing the shared resource as in \cite{6560391, Radhakrishnan201667} might mediate or negotiate between conflicting interests. Note that this assignment of a dedicated agent may be a source of contention in scenarios where multiple stakeholders are involved and compete. 

For handling scarce resources, the fields of Mechanism Design and, more generally, Game Theory apply. Notably, recent work \cite{EC16-Albert} documents a robust mechanism that allows the mechanism designer to incorporate imprecise estimates of the distribution over bidder (agents) valuations (cost functions) providing strong guarantees that the mechanism will perform at least as well as "ex-post" mechanisms, while in many cases performing better. Exploiting this trait for predictive control in buildings with multiple agents appears as promising and should be investigated.

\emph{RQ 4: Depending on the involved agents' assertiveness and willingness to cooperate, how can conflicts of control due to systemic interdependencies or resource scarcities be appropriately mitigated in multi-agent building control systems?}

\emph{RQ 5: Do agents learning to predict situations of conflict increase building energy efficiency?}

\subsection{Energy Disaggregation}
To apply predictive control, the works listed in Table \ref{tab:energyStudiesLong} require system level or appliance level sensing and metering. While sensing  devices become cheaper, professional metering installations are still expensive - especially when the effort to install is complex, e.g.~for heating distribution pipes that require welding, emptying of the pipes, and changes in the thermal insulation. The costs of installing and commissioning sub-metering can threaten the economics of projects aiming to implement any of the methods surveyed. Considering that existing buildings' BMS typically have access to main meters, several important sub-meters, and operational data, it is appealing to consider inferring appliance or system energy consumption from BMS data, i.e.~from building system operation. 

Many recent approaches to energy disaggregation (or non-intrusive load monitoring, NILM) focus on household appliance disaggregation, not large facilities. For example, \cite{Parson20141} proposes a two-stage method that generalizes to previously unseen households after having trained the appliances' probabilistic Hidden Markov Models. Similarly, \cite{Cominola2017331} combines Hidden Markov Models with time warping techniques to disaggregate household appliance data. These works have in common that they rely on a supervised training phase that could be challenged: Considering the availability of BMS operational data, possibly no calibration would be needed for predictive control purposes as long as the building system (or appliance) energy can be disaggregated with sufficient accuracy. 

Unfortunately, the problem of disaggregating energy consumption of any commercial building into individual appliances poses several technical challenges.
\begin{itemize}
\item Most disaggregation research focuses on households and their appliances. 
\item A typical commercial facility has a large number of appliances which often run simultaneously. As a consequence of this, multiple energy signatures overlap complicating the task of identifying individual appliances' signatures from the aggregate consumption. 
\item Many approaches base on the assumption that only one appliance switches its operating state at a time. This may hold for buildings with low numbers of appliances and high-frequency data, but this assumption does not hold in the case of commercial buildings with larger numbers of appliances and systems.
\end{itemize}

Furthermore, the disaggregation works in the literature focus almost exclusively on learning electricity consumption models of the different appliances typically encountered in households based on a training dataset. The models are then applied to new homes. This approach stems from the absence of historical data in a typical household. While not harmful, this transferability is not required in buildings with existing instrumentation as data is typically logged for some time. 
Thermal systems such as hydronic space heating are by definition slower than electric systems and, to the best of our knowledge, not covered at all in recent energy disaggregation literature. Only \cite{Cohn2010} addresses gas consumption disaggregation by introducing an additional sensor to analyze the acoustic response of houses' gas regulators. Given that building management systems already provides rich data about building system operations, it should be investigated if disaggregation can also do without additional equipment installations. 
Recent works attempt to disaggregate based on low-frequency data \cite{Parson20141, AAAIW1715097}. They rely on multi-second to multi-minute data, as opposed to the sub-second or even kHz data sampling rates commonly encountered in literature. Further investigation of the applicability to controlling a larger number of appliances and systems is required though. 

\emph{RQ 6: Are recent disaggregation approaches sufficient to enable predictive control of large buildings with existing BMS installations?}

\emph{RQ 7: Can the disaggregation approaches be extended beyond electricity or are different methods required?}

\section{Conclusion}\label{sec:conclusion}

The survey review in Sections \ref{sec:back} and \ref{Sec:BuildingEnergyPerformance} leads to the following insights:

\begin{itemize}
\item The majority of studies focuses on optimizing operational energy cost or consumption. Most take into account thermal comfort, often by formulating a multi-objective optimization, e.g.~by calculating the Pareto front. Little attention is given to exergy or alternative KPIs such as system performance (UPR), green factor, or CO$_2$ savings. Suffice to say, studies performing optimization of energy costs may result in higher building energy usage.

\item Where PMV and PPD are used for thermal comfort assessment, works do not discuss whether they use the static or adaptive definitions of the comfort KPI. 

\item A minority of studies addresses the aspect of BMS integration. 

\item Systemic interdependencies or conflicting control commands are rarely studied. 

\item Most works perform validation in the form of simulation. Fewer studies validate the control in experiments - and of those that do, the majority uses relatively short experimental periods. 

\item Few works account for weather conditions when assessing energy performance. That may be appropriate for simulations as conditions are reproducible. However, real world experiments' results may be influenced by changing weather conditions. Even fewer works apply statistical methods to draw robust conclusions by quantifying the confidence intervals of the savings they achieve.

\item Most works do not explain how the features taken into account are derived or why they are used.  

\item Nature-inspired optimization heuristics such as the Genetic Algorithm or Particle Swarm Optimization are frequently encountered. Robust optimization techniques are not prominent in the field of data-driven predictive building control. 

\item Data-driven predictive control studies reach similar effect sizes as simulation-driven MPC studies. Neural Networks, and to a lesser extent Reinforcement Learning, are commonly used in the literature.

\item Although the field of predictive building control is clearly related to the CPS concept, none of the works explicitly applies the general CPS research findings. In particular, a methodology for evolving buildings into energy efficient predictive CPS is lacking.
\end{itemize}

With these insights in mind, future research is guided by the following questions derived in Section \ref{sec:open}. Studies should investigate each of the following research questions:

\emph{RQ 1: What is a suitable methodology to evolve existing buildings into a CPS for higher levels of operational efficiency?}

\emph{RQ 2: How to perform semantic feature selection for predictive building control?}

\emph{RQ 3: How to transfer a particular building’s predictive control strategies to another building?}

\emph{RQ 4: Depending on the involved agents' assertiveness and willingness to cooperate, how can conflicts of control due to systemic interdependencies or resource scarcities be appropriately mitigated in multi-agent building control systems?}

\emph{RQ 5: Do agents learning to predict situations of conflict increase building energy efficiency?}

\emph{RQ 6: Are recent disaggregation approaches sufficient to enable predictive control of large buildings with existing BMS installations?}

\emph{RQ 7: Can the existing disaggregation approaches be extended beyond electricity or are different methods required?}